\UseRawInputEncoding
\documentclass[12pt]{article}

\usepackage{graphicx}
\usepackage{amsmath,amssymb}
\usepackage{mathtools,xcolor}
\usepackage{float}
\usepackage{authblk}
\usepackage{bm}
\newcommand{\tr}{\operatorname{Tr}}

\newcommand{\rmd}{\mathrm{d}}

\newcommand{\rmF}{\mathrm{F}}

\newcommand{\rme}{\mathrm{e}}
\newcommand{\rmi}{\mathrm{i}}

\newcommand{\braket}[2]{\langle #1 | #2 \rangle}
\newcommand{\ketbra}[2]{| #1 \rangle\langle #2 |}

\newcommand{\la}{\langle}
\newcommand{\ra}{\rangle}

\newcommand{\be}{\begin{equation}}
\newcommand{\ee}{\end{equation}}
\newcommand{\ba}{\begin{eqnarray}}
\newcommand{\ea}{\end{eqnarray}}

\def\h{\hskip 1cm}

\def\lo{\longrightarrow}
\def\a{\alpha}
\def\b{\beta}

\newtheorem{remark}{Remark}

\title{\bf GHZ states as near-optimal states  for reference frame alignment}

\author{Mear M. R. Koochakie\footnote{email: m.koochakie@gmail.com}}
\author{Vahid~Jannesary\footnote{email: vahid.jannesary@gmail.com}}
\author{Vahid~Karimipour\footnote{email: vahid@sharif.edu}}

\affil{\it Department of Physics, Sharif University of Technology, Tehran 14588, Iran}

\date{}

\begin{document}

\maketitle

\begin{abstract}
	Let two coordinate systems, in possession of Alice and Bob, be related to each other by an unknown rotation $R\in SO(3)$.  Alice is to send identical states $|\psi_0\ra$ to Bob who will make measurements on the received state and will determine the rotation $R$. The task of Bob is to estimate these parameters of the rotation $R$  by the best possible measurements.  Based on the Quantum Fisher Information, we show that  Greenberger-Horne-Zeilinger (GHZ) states are near optimal states for this task. Compared to the optimal states proposed before, the advantage of $GHZ$ states are that they can be more easily prepared experimentally, and more importantly, we show  concrete measurements which will allow Bob to determine the rotation $R$.   We also study the robustness of these states in keeping their encoded information, against common sources of noises.  
\end{abstract}
	
\section{Introduction}
A class of quantum estimation theory problem can be stated as follows: A unitary operator $U(\theta)$ acts on a quantum state $|\psi_0\ra$ turning it into $|\psi(\theta)\ra=U(\theta)|
\psi_0\ra$. The parameter $\theta$ is for example related to the coupling constant of a certain Hamiltonian, i.e. the strength of a magnetic field whose direction is known to us and our task is to estimate its strength by performing the best measurements on the state $|\psi(\theta)\ra$. An important question is which reference state encodes this information in an optimal way, where  "optimality" can be defined in various ways. Intuitively one can say the optimal state should be the one which is the most sensitive to variations of $\theta$. For example, if the direction of the magnetic field is known to be in the $z-$ direction and $\theta$ encodes only the strength of this magnetic field, then the optimal states are those which are in the $xy$ plane, i.e. the so-called equatorial states, and the more these states are tilted toward the z-axis, the less effective they are in carrying out the information. In passing, we note that it has been shown that in this simple setting, entanglement can enhance the precision of parameter estimation \cite{gio, dariano}. \\

Suppose however that we do not even know the direction of the magnetic field and only a probability distribution of the directions is at hand. Then the above intuitive argument is no longer valid and we have to resort to a quantitative measure to answer this question. Besides, using tools from multi-parametric estimation theory, we should also do an averaging over all directions or parameters in order to find the optimal state. Such an optimal state will certainly be the optimal state for a certain direction but not for the others, but on the average, it will be optimal when the probability distribution is taken into account. For a review see \cite{paris, paris2}.\\

In a more general setting the problem can be stated as follows: Let $G$ be a group of transformations and $|\psi\ra$ a reference state. The information of a group element $g=g(\theta_1,\cdots \theta_n)\in G$ is encoded into a unitary representation $U(g)$ acting on the state $|\psi\ra$ in the form $|\psi(g)\ra=U(g)|\psi_0\ra$ and the parameters $(\theta_1,\cdots, \theta_n)$ of $g$ are to be estimated by making suitable measurements on this state. Given that the elements of the group are chosen according to a measure 
$P(g)dg$ with $\int_G P(g)dg=1$, our task is to find the optimal reference state $|\psi\ra$ for this task. \\ 

In a particular setting, $G$ can be the group of rotations $SO(3)$ and the parameters $(\theta_1,\theta_2,\theta_3)$ can be the Euler angles. In this setting this is relevant among other things to the problem of reference frame alignment, i.e. when two previously aligned reference frames are distorted due to noise.  While classical means are always available for reference frame alignment, it has become apparent in recent years that quantum resources like entanglement, can lead to the improved Heisenberg scaling in the achievable precision. \cite{ gio, paris2, dariano}. Moreover there are other applications closely related to this problem,  like estimating the direction and magnitude of a magnetic field, or the amount of distortion of photon polarization sent through an optical fiber.

\subsection{Two main approaches to the problem of shared reference frames}
Before proceeding we should discriminate between two approaches which has been followed in the past years for dealing with the problem of reference frames. In one approach which is the subject of this paper, it is tried to align the two reference frames by quantum mechanical means before doing any quantum information task. By quantum mechanical means we means that one party sends an optimal  reference to the other which is to be measured and based on the results of the measurement the other reference frame should be properly rotated to be aligned with the first one. This is the approach which was motivated by \cite{gispop, mass} and initiated in \cite{peres1, peres2}. This approach was then developed in several works including \cite{reza, bagan, chir, kolen, gold}. There is however another approach which seeks to do quantum information task without aligning the reference frames. In this approach it is either tried to encode classical or quantum information into particular states which are insensitive to alignment \cite{Rudolph1, Rudolph2, rezaqkd, Beheshti} or else try to model  the effect of this misalignment as a noisy channel and assess their effect on the classical or quantum information which is being communicated \cite{ahmadi1, ahmadi2, ahmadi3}.   A general theory of reference frames and its relation with asymmetry and resource theory has also been developed in \cite{MarvianSpekkens1, MarvianSpekkens2, MarvianSpekkens3}.\\

\subsection{A brief history of the first approach}

The approach that we follow here is the first one where the two parties try to align their reference frame by estimation of optimal states sent transferred between the frames. This approach has a long history for itself which is quite different from the second one. First  Gisin and Popescu showed in \cite{gispop} that sending two anti-parallel spin one-half particles is better than two parallel spins for sharing a direction, then Massar and Popescu \cite{mass} showed that the fidelity $F$ of transmission of a direction with $N$ parallel spins scales as $1-F\sim \frac{1}{N}$ when the receiver (Bob) is allowed to make collective muti-qubit measurements.  Later,  Peres and Scudo \cite{peres1} and Bagan et al \cite{bagan} showed independently that the fidelity can increase quadratically (i.e. $1-F\sim \frac{1}{N^2}$) with the number of particles if the state is encoded into an entangled state of $N$ spin one-half particles, namely a state which is an eigenstate of total spin operator in the desired direction. Alleviating the need for multi-qubit measurements, it was shown in  \cite{reza} that one can do single qubit measurements on singlet states shared between the two parties and use the correlations in measurements to set up a common direction with the same fidelity as in Massar and Popescu \cite{mass}. The figure of merit used in all these works was the fidelity between the actual direction sent by Alice and the direction guessed by Bob, averaged over all sent directions.\\

These studies naturally were extended to sending copies of a specific state which encodes the information of a  whole Cartesian frame \cite{peres2,bagan,chir,kolen,gold}. The idea is that Alice takes a particular state  $|\psi_0\ra$ in her own Caterzian frame $(x,y,z)$ and sends copies of it to Bob whose frame $(x',y',z')$ is not aligned with that of Alice. This state appears in the frame of Bob as 
\be
|\psi(\bm\theta)\ra=U(\bm\theta)|
\psi_0\ra ,
\ee
where  $\bm\theta=(\theta_1,\theta_2,\theta_3)$ are the three parameters which designates the rotation operator $U(\bm \theta)$ aligning Bob's frame with that of Alice. It is important to note that $|\psi_0\ra$ itself does not have any intrinsic dependence on $\bm\theta$. Bob then performs optimal measurements on this state in his own frame to estimate the values of the parameter $\bm\theta$.   The aim of optimal measurements of Bob is to estimate the values of these parameters from the statistics of his measurements.
In \cite{peres2}, it was shown that 
despite the spherically symmetric potential of a Hydrogen atom a particular Rydberg state in the $n-th$ level of a Hydrogen like atom, $|\psi_0\ra=\sum_{j=0}^{n-1}\sum_{m=-j}^{j}a_{jm}|j,m\ra$, can indeed optimally encode this information. Here the figure of merit used was the average fidelity over all orientations of frames and the explicit form of the states, i.e. the values of $a_{j,m}$, were found numerically. \\

Besides states of Rydberg atoms, collective spin states of a groups of $N$ spin 1/2 particles have also been studied in a number of works and this is the approach that we will also follow in our work. 
That is the state we consider is  of the type 
\be
|\psi_0\ra=\sum_{i_1,\cdots i_N} \psi_{i_1,i_2,\cdots i_n}|i_1, i_2, \cdots i_N\ra
\ee
where $|i_k\ra\in \{|0\ra, |1\ra\}$ are the basis spin states of one particle. 
In \cite{kolen}, the multi-parameter Cram\'{e}r Rao bound was considered and it was shown that when the frames are slightly mis-aligned (an approach called the local approach in \cite{kolen}) simpler states can encode the information of Alice's frame. The basic premise in the local approach of \cite{kolen} is that for small misalignment, the rotation operator is linearized

\be\label{lin} U(\bm\theta)\approx I+i\bm\theta\cdot{\bf S},\ee
which leads to a drastic simplification of the Quantum Fisher Information matrix. 
Here ${\bf S}=(S_x, S_y, S_z) $ is the total spin operator acting on the spins, i.e. 
\be
S_a=\frac{1}{2}\sum_k  \sigma_{a,k}.
\ee      
The optimal states are thus found to be the so-called anti-coherent states of spin $j-$ systems, i.e. those which satisfy the following conditions
\be\label{cond}
\la \psi_0|S_a|\psi_0\ra=0\h \la\psi_0| S_a^2|\psi_0\ra=\frac{j(j+1)}{3}\h a=x,y,z.
\ee
Therefore the authors of \cite{kolen} found analytical forms for the optimal states which turned out to be certain specific states of $N$ particles, studied in \cite{zimba} and named anti-coherent states. When represented in terms of Majorana representation \cite{maj,zimba}, they correspond to Platonic solids:  Any symmetric state can be decomposed as  superposition of product states, i.e. 
\be
|\psi\ra=N\sum_{\sigma\in S_N}|\vec{n}_{\sigma(1)}\ra \otimes |\vec{n}_{\sigma(2)}\ra \otimes\cdots |\vec{n}_{\sigma(N)}\ra,
\ee
where $N$ is a normalization constant, the summation is performed over all permutations and the states 
$	|\vec{n}_{k}\ra$ are pure states on the Bloch sphere. There is a one-to-one correspondence between any such state and the points on the surface of Bloch sphere \cite{zimba}.\\

Quite recently \cite{gold} the problem has been studied in the so called global approach, i.e. where the two frames are misaligned by arbitrary finite rotations. To calculate the Fisher information and find the optimal state in this approach, it is necessary to parameterize  the rotation operator by the three Euler angles, as  
\be\label{Euler} U(\a,\beta,\gamma)=e^{-i\a S_z}e^{-i\beta S_y}e^{-i\gamma S_z}.\ee
Intriguingly and despite the asymmetry of the three axes, what the authors of \cite{gold}  obtain are still regular polyhedrons. They also obtained new forms of symmetric 
states corresponding to Archimedean solids which contain the states of \cite{kolen} as a subset, figure (\ref{plane2}). \\

The basic premise in the work of Goldberg and James \cite{gold} which leads to this symmetry of optimal states, despite the asymmetry of their parameterization is that they use a result of \cite{kolen}, namely (\ref{cond}) which is valid only for symmetric parameterization (\ref{lin}) of the rotation operator. \\

\noindent {\bf Remark:} It is crucially important to note that in all of these works the optimal state cannot be prepared experimentally due to its complicated form, whether it be calculated analytically or numerically, whether in the local or global approach. \\

\subsection{Our main results}
In this article we start from scratch, use Euler parameterization for rotation and show that the optimal state is in fact a GHZ state of the form 
\begin{equation}
|\psi_{0}\ra=	\frac{1}{\sqrt{2}}(|\frac{N}{2},\frac{N}{2}\ra+|\frac{N}{2},-\frac{N}{2}\ra)=\frac{1}{\sqrt{2}}(|\uparrow, \uparrow,\cdots \uparrow\ra+|\downarrow, \downarrow,\cdots \downarrow\ra),
\end{equation}
where the states $|\uparrow\ra$ and $|\downarrow\ra$ are the spin up and down in Alice's $z-$ direction and $\delta$ is an arbitrary phase. (In fact the two terms can also have a phase difference $e^{i\delta}$ without altering any of the results.) The basic point is that when a rotation operator is decomposed as in Eq. (\ref{Euler}), the sensitivity of the sent state to variations of the three parameters is certainly not equal and there is an obvious difference between the z-direction and the other directions. Therefore it is understandable that the optimal state may not correspond to a regular polyhedron. As we will see, the optimal state that we will obtain is still a symmetric state with a nice Majorana representation shown in figure (\ref{plane2}). Had the Euler decomposition been chosen in a different form, i.e. with different orders of rotation, the GHZ state would have been different. In any case, the important point is that such states can be created quite easily in the laboratory with a quantum circuit. Our analysis shows in yet another quantum information task,  the usefulness of GHZ states as superposition of macroscopically distinct states. \\

The figure of merit that we use is the related to Quantum Fisher Information of the encoding state uniformly averaged over all rotations. Based on the this figure of merit, we show that  Greenberger-Horne-Zeilinger states are optimal states for this task. The advantage of these states are that, compared to the optimal states which have been proposed so far,  they can be experimentally prepared \cite{ghzexp1,ghzexp2, ghzexp3, ghzexp4} . We also study the robustness of these states in keeping their encoded information, against common sources of noises. \\

The structure of this paper is as follows: In section (\ref{crb})  we set up the problem and explain in detail the figure of merit which leads to the optimal state. Then we will show that the GHZ states are near optimal in a sense which will be explained. Finally we compare our final results with those of reference \cite{gold}.   We conclude the paper with a conclusion. \\

	\section{Cram\'{e}r-Rao bound for frame alignment }\label{crb}   
	   As explained in the introduction,  Alice takes a particular state  $|\psi_0\ra$ in her own Caterzian frame $(x,y,z)$ and sends copies of it to Bob whose frame $(x',y,z')$ is not aligned with that of Alice. This state appears in the frame of Bob as 
	  $
	  |\psi(\bm\theta)\ra=U(\bm\theta)|
	  \psi_0\ra
	  $
	  where  $\bm\theta=(\theta_1,\theta_2,\theta_3)$ are the three parameters which designates the rotation operator $U(\bm \theta)$ aligning Bob's frame with that of Alice. The state  $|\psi_0\ra$ itself does not have any intrinsic dependence on $\bm\theta$. 	
	  	  Let us denote the estimated values by $\bm\theta_e=(\theta_{1,e},\theta_{2,e},\theta_{3,e})$. Here optimality of the whole process of estimation depends on the state $|\psi_0\ra$ chosen by Alice and the clever measurements of Bob. In all the works which have taken the Quantum Fisher Information as their figure of merit, as also in the present work, the emphasis is to find the optimal state $|\psi_{0,opt}\ra$ which Alice should send, rather than on the optimal measurement performed by Bob and the estimation method he may use. These may be quite complicated and varied.  What really allows these two optimal choices to be made separately is the celebrated mutli-parameter Cram\'{e}r Rao bound, which  states that the variance between the real values of the parameters $\theta_i$ and their estimated values $\theta_{i,e}$ satisfies the following matrix inequality \cite{cr1,cr2,hels,petz}
	  	   	\be\label{crb-ineq}
	  	   \Delta(\bm\theta_e,\bm\theta)\geq F^{-1}(\psi(\bm\theta)),
	  	   \ee
	  	   where 
	  	   \be
	  	   \Delta(\bm\theta_e,\bm\theta)_{m,n}=\la (\theta_{m,e}-\theta_m)(\theta_{n,e}-\theta_n) \ra ,
	  	   \ee
	  	   and
	  	   \ba\label{fij}
	  	   F_{mn}(\psi(\bm\theta)) &=& 2 \braket{\partial_{m}\psi(\bm\theta)}{\partial_{n}\psi(\bm\theta)} + 2 \braket{\partial_{n}\psi(\bm\theta)}{\partial_{m}\psi(\bm\theta)}\nonumber \\
	  	   &+& 4 \braket{\psi(\bm\theta)}{\partial_{m}\psi(\bm\theta)} \braket{\psi(\bm\theta)}{\partial_{n}\psi(\bm\theta)}.
	  	   \ea
Here $|\psi(\bm\theta)\ra$ is the state $|\psi_0\ra$ as appeared in the frame of Bob, that is $|\psi(\bm \theta)\ra=U(\bm\theta)|\psi_0\ra$. The important point about (\ref{crb-ineq}) is that the right hand side is independent of the measurement process and any estimation procedure that Bob may use. It sets a theoretical lower bound on the variance of the values of estimated parameters compared with their real values. 
	  	    Note also that we have used the form of Quantum Fisher Information for pure states, since we have used the fact that $F$ is a convex function of the parameterized quantum states, and hence takes its optimum value at the extreme points of the space of states, that is the pure states.  \\
	  	   
An important question is whether this lower bound can be saturated and it has been shown in \cite{matsu} that  this is indeed the case if 
	  	   \be
	  	   Im\la \partial_m\psi(\bm\theta)|\partial_n\psi(\bm\theta) \ra=0.\ee
	  	   	  	   One usually defines a cost function as the sum of variances of all three parameters, i.e. 
	  	   	  	   \be
	  	   C(\bm\theta_e,\bm\theta):=\sum_{m}\la (\theta_{m,e}-\theta_m)^2 \ra =Tr\Delta(\bm\theta_e,\bm\theta),
	  	   \ee
	  	   which is bounded below as 
	  	   \be
	  	    C(\bm\theta_e,\bm\theta)
\geq Tr\big[F^{-1}(\psi(\bm\theta))\big]=Tr\big[ F^{-1}(U(\bm\theta)\psi_0)\big].	  	   \ee
	  	   	  	   
	  	  \noindent Therefore for communicating a "{\it{known}}" parameter $\bm\theta$, the optimal state $|\psi_{0,opt}\ra$ is the state which minimizes the right hand side. Naturally this state depends on the parameter $\bm\theta$, therefore we can write $\psi_{0,opt}=\psi_{0,opt}(\bm\theta)$.\\

While these relations are valid for any process in which a set of "{\it{known}}" parameters are encoded into quantum states and communicated to some other point, in the problem of frame alignment, we are faced with a situation in which  the parameter is not known to the sender. Therefore in this case the optimal state cannot have any dependence on $\bm\theta$ even after optimization. Instead one has to define a cost function averaged over all values of the parameter $\bm \theta$. A natural cost function is defined as follows 
	  	   \be
	  	   C:=\int d\bm\theta Tr(\Delta(\bm\theta_e,\bm\theta)),
	  	   \ee  	   
	  	    where $d\bm\theta$ is a suitable measure over the space of parameters. Thus $C$ depends only on the estimation procedure and the initial state $|\psi_0\ra$, both of which are suppressed in the notation for $C$ for simplicity.   Then equation (\ref{crb-ineq}) leads to 
	  	      
	  	      \be\label{crbav}
	  	      C\geq \int d\bm \theta \ Tr(F^{-1}(U(\bm\theta)\psi_0)=: Tr(\overline{F^{-1}}).
	  	      \ee
	  	      One can obtain a still lower bound \cite{gill1,gill2} by adopting the so called van Trees inequality and the Baeysian-Cramer Rao bound, however we suffice here to this slightly looser bound. 
	  	    In view of the convexity of the quantum Fisher matrix \cite{conv} and the inequality $Tr(X^{-1})Tr(X)\geq n^2$ for any $n$ dimensional positive matrix, one can also write
	  	      \be\label{crbavin}
	  	    C\geq \frac{9}{\int d\bm \theta \ Tr(F(U(\bm\theta)\psi_0)}=:\frac{9}{Tr(\overline{F})}.
	  	    \ee
	  	    
	  	   \noindent {\bf Remark:} 
	  	     	 While inequality (\ref{crbav}) can be saturated if for each $\bm\theta$, one inserts $\psi_{0,opt}(\bm\theta)$ inside the integral on the right hand side, the inequality in (\ref{crbav}), is not normally saturated, that is, the quantum Fisher information, as used in (\ref{crbav}) and (\ref{crbavin})  gives a slightly less tight lower bound compared to the case where a known parameter is to be communicated by encoding it into a quantum state. Therefore no matter how the right hand side of (\ref{crbav}) is minimized, the Cram\'{e}r-Rao bound cannot be saturated for frame alignment and all the optimal states found in different works, including the present one,  should actually be called near-optimal states.  \\
	  	     	 
What we will do is to calculate the quantum Fisher information matrix and show that the GHZ state which almost maximizes the quantity $Tr\overline{F}$.  The large benefit which comes from this approach is the simplicity of creating a GHZ state in the laboratory compared with other states.

	\section{The Quantum Fisher Information for frame alignment}\label{sec-generalCase}

\noindent  The state $|\psi_0\ra$ that Alice sends appears in  Bob frame as  \be |\psi({\bm\theta})\ra=U({\bm\theta})|\psi_0\ra,\ee  where 
$U({\bm\theta})$ is the rotation which align the two coordinate systems. There are a multitude of ways for parameterization of a rotation operator. The parameterization $U({\bm\theta})=e^{i\bm\theta\cdot{\bf S}}$, which has many good properties is suitable only for small rotations, when it comes to calculating the Fisher matrix. Instead we use the Euler angle parameterization which allows us to determine the Fisher matrix for finite rotations. Thus we use 
\begin{equation}\label{Upar}
U(\bm\theta) = U(\a,\b,\gamma)=\rme^{-\rmi \alpha S_z} \rme^{-\rmi \beta S_y} \rme^{-\rmi \gamma S_z}\equiv Z(\alpha) Y(\beta) Z(\gamma),
\end{equation}
where $\a, \beta$ and $\gamma$ are the Euler angles which transform Alice's frame to that of Bob.  It should be noted that there are many variants of the Euler angles and there is no preference among them, since they all lead to valid lower bounds for precision. In other words, had we decomposed the rotation in a different form, we would have ended up again  with an optimal state as a GHZ state in a different direction, but this is only a matter of labeling qubit states. The fact that Alice should send a GHZ state remains unchanged. More importantly, when Bob does measurements, the way this rotation has been parameterized is not relevant, see section (\ref{MES}). \\

	\noindent We can now use Eq.~\eqref{fij} and calculate QFI matrix of $|\psi(\alpha,\beta,\gamma)\rangle=U(\a,\b,\gamma)|\psi_0\ra$. 	
For a pure state, the components of the Quantum Fisher Information (QFI) are given by \cite{Liu2014, Liu2015}

\begin{equation}\label{eq:QFIpure}
\rmF_{mn} =   4\mathrm{cov}\left(H_{m},H_{n}\right),
\end{equation}
where $H_{m}=\rmi(\partial_{m}U^{\dagger})U,\  \ m=\a, \beta, \gamma $  and
\begin{equation}\label{eq:cov(H_m,H_n)}
\mathrm{cov}(H_{m},H_{n}):=
\frac{1}{2}\langle\psi_{0}|\{H_{m},H_{n}\}
|\psi_{0}\rangle-\langle\psi_{0}|H_{m}|\psi_{0}\rangle\langle\psi_{0}|
H_{n}|\psi_{0}\rangle.
\end{equation}

	\noindent 
	It is straightforward to check that
	\begin{align}
	H_\a &= -\cos \beta S_z + \sin \beta (\cos \gamma S_x - \sin \gamma S_y), \label{eq:H_a} \\ 
	H_\beta&=-\cos \gamma S_y - \sin \gamma S_x,\label{eq:H_beta}\\ 
	H_\gamma &= - S_z.\label{eq:H_gamma} 
	\end{align}
In particular we have 
\be
F_{m,m}=4\la \psi_0|H_m^2|\psi_0\ra-4\la\psi_0| H_m|\psi_0\ra^2,\ \ m=\a, \beta, \gamma.
\ee

\noindent	  To calculate $\overline{\tr F}$,
	we need a prior distribution for the parameters $\alpha, \beta$ and $\gamma$. We take a uniform distribution (SO(3) Haar measure)
	
	\begin{equation}\label{eq:distEulerParam}
	z(\alpha,\beta,\gamma)\, \rmd\alpha \rmd\beta \rmd\gamma= \frac{1}{8 \pi^2} \sin \beta \,\rmd\alpha \rmd\beta \rmd\gamma.
	\end{equation}
	Here we have assumed that the distribution of the $z$ axis is uniform over a sphere in Bob's frame. This gives the factor $\frac{1}{4\pi}\sin\beta \,\rmd\alpha \rmd\beta$ in (\ref{eq:distEulerParam}). Once the $z$ or the $x$-$y$ plane is fixed, the $x$ axis can be uniformly distributed in this place, with a distribution $\frac{1}{2\pi}\,\rmd\gamma$, hence Eq.~\eqref{eq:distEulerParam}. \\
	
	\noindent Using the above distribution, we show that (see the appendix)
		\begin{eqnarray}\label{eq:mtqEuler}
			Tr(\overline{F}) &=& \frac{16}{3} (\Delta S_z)^2 + \frac{10}{3} (\Delta S_x)^2 + \frac{10}{3} (\Delta S_y)^2 \cr
			&=&\frac{16}{3} (\la S_z^2\ra-\la S_z\ra^2) + \frac{10}{3} (\la S_x^2\ra-\la S_x\ra^2) + \frac{10}{3} (\la S_y^2\ra-\la S_y\ra^2)\cr&=&
				\frac{10}{3} \la {\bf S}\cdot {\bf S}\ra+2\la S_z^2\ra-\la S_x\ra^2-\la S_y\ra^2-\la S_z\ra^2 .
				\end{eqnarray}
This expression clearly shows the special role which the $z-$ axis plays compared with the other two axes. Note however that  had we decomposed the rotation in a different form, we would have ended up again  with an optimal state as a GHZ state in a different direction, but this is only a matter of labeling qubit states. The fact that Alice should send a GHZ state remains unchanged. 

To find the state $|\psi_0\ra$, which maximizes this quantity, we take an arbitrary state of $N$ spin 1/2 particles,  
\be
|\psi_0\ra=\sum_{j=0,\frac{1}{2}}^{\frac{N}{2}}\sum_{m=-j}^ja_{j,m}|j,m\ra ,
\ee
where $j$ runs from 0 or $1/2$ for even and odd $N$ respectively and $|j,m\ra$ is the familiar angular momentum notation. It is then obvious that $Tr(\overline{F})$ is maximized when $j$ takes its largest value and $\la S_x\ra=\la S_y\ra=\la S_z\ra=0$. Such a state is nothing but a GHZ state, i.e. a superposition of macroscopically distinct states

	\noindent	
	\begin{equation}\label{ghzn0}
	| \psi_0 \rangle = |GHZ\ra=\frac{|\frac{N}{2},\frac{N}{2}\rangle + e^{i\delta} |\frac{N}{2},-\frac{N}{2}\rangle}{\sqrt{2}},
	\end{equation}
	where $\delta$ is an arbitrary phase and $|\frac{N}{2},\frac{N}{2}\rangle=|\uparrow, \uparrow, \cdots \uparrow\ra$ and $|\frac{N}{2},-\frac{N}{2}\rangle=|\downarrow, \downarrow, \cdots \downarrow\ra$ in which the spin up and spin down are along  the $z$ directions of Alice frame. 
These states maximize $Tr(\overline{F})$ to the value $
	\frac{N(4N+5)}{3}. 
	$ It turns out that the phase $\delta$ has no effect on the QFI and hence we set $\delta=0$ hereafter. \\

\subsection{Measurement}\label{MES}
\begin{figure}[H]
	\centering
	\includegraphics[width=1.0\linewidth]{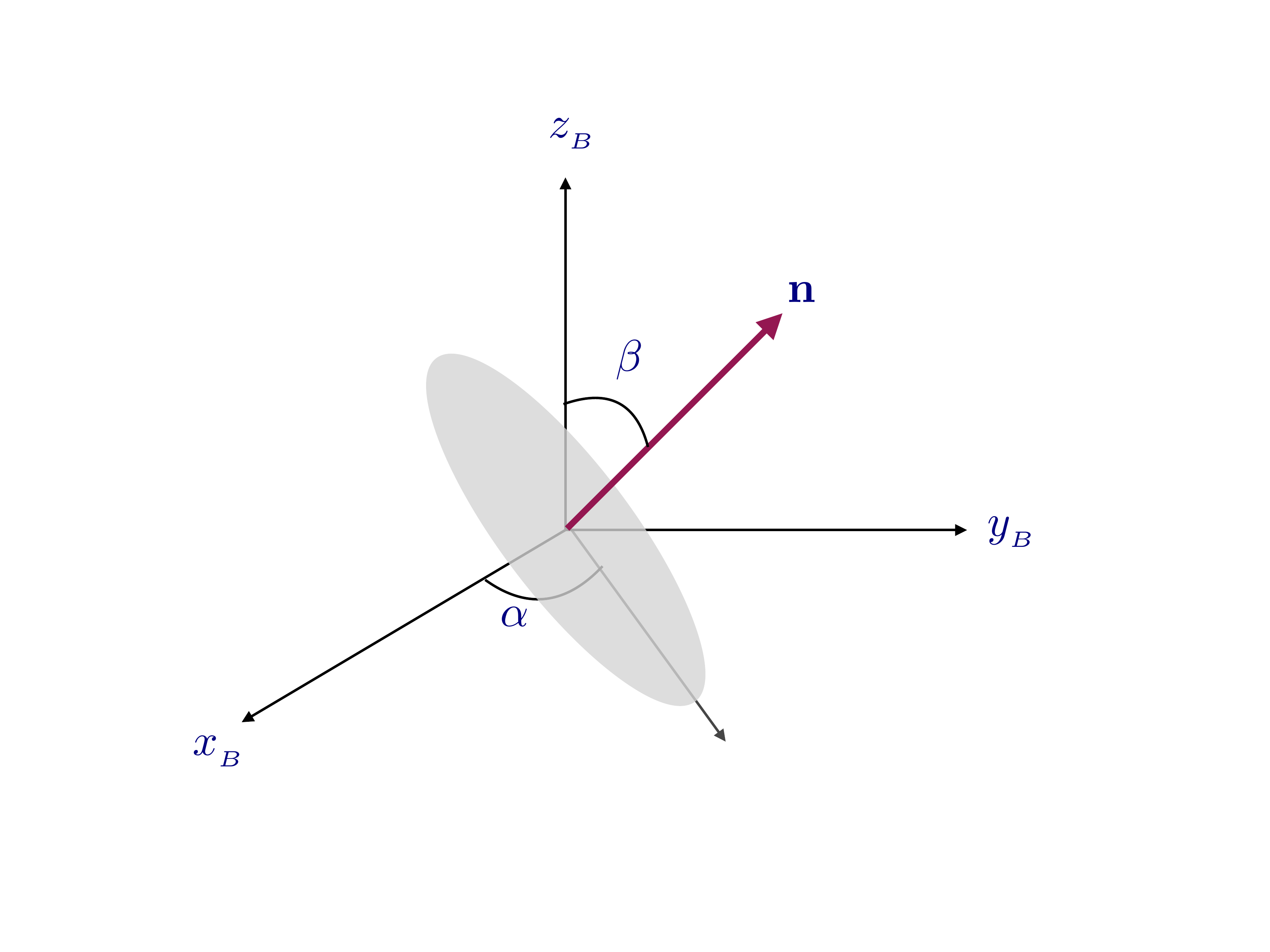}\vspace{-2cm}
	\caption{The GHZ state $|\psi_0\ra=\frac{1}{\sqrt{2}}(|z\ra^{\otimes N}+|-z\ra^{\otimes N})$ sent by Alice, appears as $|\psi'_0\ra=\frac{1}{\sqrt{2}}(|\bf n\ra^{\otimes N}+e^{i\gamma N}|-\bf n\ra^{\otimes N})$ in frame of Bob. Bob can then determine the angles $\a\ ,\beta$ and $\gamma$ by suitable measurements as detailed in section (\ref{MES}).  }
	\label{zazb}
\end{figure}

 In general it is not easy to determine the optimal measurement which saturates the Cramer-Rao bound. To the best of our knowledge, simple measurements have not been proposed  for anti-coherent symmetric states proposed in \cite{kolen} and \cite{gold} or for the Rydberg states proposed in \cite{peres2}.  Nevertheless it is a highly legitimate question as to how after all , Bob determines the coordinate system of Alice by doing measurements on a single type state that Alice sends him. Here we propose a  measurement procedure that he can do for this task, without claiming that it is the optimal one. Note that if the two frames are related by the rotation $U(\bm\theta)^\dagger$, the state $|\psi_0\ra$ will appear in Bob frame as $U(\bm\theta)|\psi_0\ra$, where $U(\bm\theta)$ is given in (\ref{Upar}) . Therefore the GHZ state $|\psi_0\ra=\frac{1}{\sqrt{2}}(|z\ra^{\otimes N}+|(-z)\ra^{\otimes N})$ appears in Bob's frame as 
   $|\psi'_0\ra=\frac{1}{\sqrt{2}}(|{\bf n}\ra^{\otimes N}+e^{-i\gamma N}|-{\bf n}\ra^{\otimes N})$, where
   ${\bf n}$ is the unit vector with polar coordinates $(\a,\beta)$ as shown in figure (\ref{zazb}). Bob's task is to do measurements on this state and discern $\a, \beta$ and $\gamma$. Knowing that  
    $|{\bf n}\ra$ and $|-{\bf n}\ra$ are spin states given by $|{\bf n}\ra=\left(\begin{array}{c}\cos\frac{\beta}{2}\\ \sin\frac{\beta}{2}e^{i\alpha}\end{array}\right)$ and  $|-{\bf n}\ra=\left(\begin{array}{c}-\sin\frac{\beta}{2}\\ \cos\frac{\beta}{2}e^{i\alpha}\end{array}\right)$ respectively,  Bob concludes that 
    $\la \psi'_0|S_a|\psi'_0\ra=0\ \ \forall\  a=x, y, z\ $. This means that he cannot determine $\alpha$ and $\beta$ simply by the expectation values of total spin component. However he can determine these angles by the variance of spin measurements, since a simple calculation reveals to him that 
     	\ba
	\la\psi_0| S_x^2|\psi_0\ra &=& \frac{1}{4}(N+N(N-1)\sin^2\beta \cos^2\alpha)\cr
		\la\psi_0| S_y^2|\psi_0\ra &=& \frac{1}{4}(N+N(N-1)\sin^2\beta \sin^2\alpha)\cr
			\la\psi_0| S_z^2|\psi_0\ra &=& \frac{1}{4}(N+N(N-1)\cos^2\beta).
	\ea
This  implies that by measuring the variance of any two components he can determine the two angles $\a$ and $\beta$.  It remains for him to determine the angle $\gamma$.  To this end he uses a second bunch of states $|\psi'_0\ra$ and rotate them to align them along his $z$ axis in which case their form will be given by 
\be
|\psi'_0\ra=\frac{1}{\sqrt{2}}(|0\ra^{\otimes N}+e^{-i\gamma N}|1\ra^{\otimes N})
\ee
where for convenience, we have used the computational notation, and he measures the operator $\Pi=\sigma_{x,1}\sigma_{x,2}\cdots \sigma_{x,N}$. He simply finds
\be
\la \psi'_0|\Pi|\psi'_0\ra=\cos\gamma N
\ee
from which he determines $\gamma$. 
\subsection{Comparison}
\begin{figure}[!ht]
	 	\centering
	 	\includegraphics[width=1.0\linewidth]{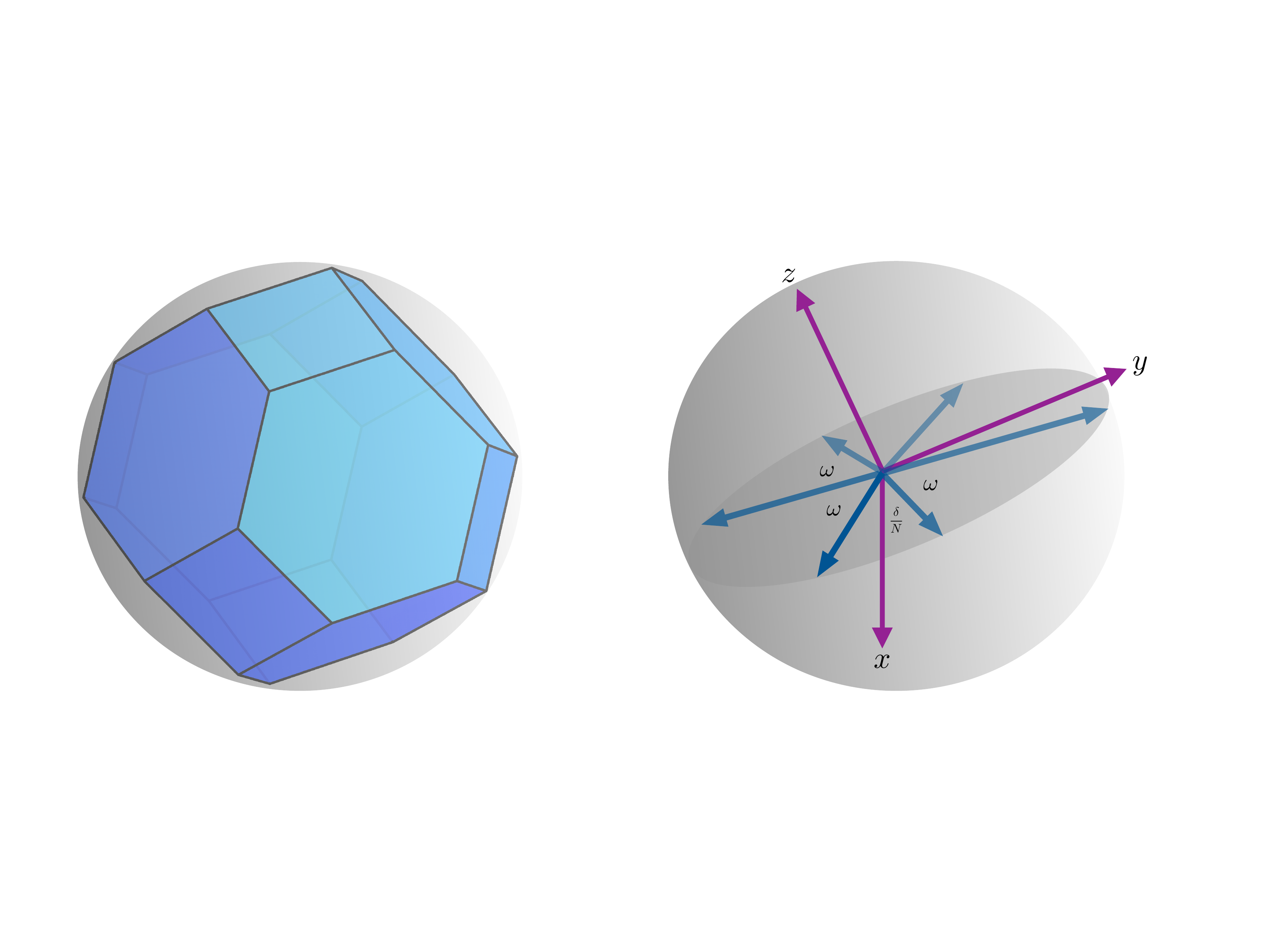}\vspace{-2cm}
	 	\caption{Left: Majorana representation of an optimal states corresponding to an Archimedean solid,  \cite{gold}. Each vertex corresponds to a pure state on the Bloch sphere and the full state is a symmetric superposition of these states. Right: Majorana representation of the GHZ state (\ref{ghzn0}) for communicating a frame, when the rotation is parameterized in terms of Euler angles.} 
	 	\label{plane2}
	 \end{figure}

As can be seen, in  $\tr(\overline{F})$ the role of $z$ direction differs from other directions.
	This is due to the way Euler rotations are parameterized and is also reflected in the Majorana representation of the state (\ref{ghzn0}), shown in figure (\ref{plane2}) which is a superposition of equally spaced product states in the equatorial plane of Alice, namely 
	
	\be
	|\psi_0\ra=N\sum_{\sigma}|z_{\sigma(1)}\ra|z_{\sigma(2)}\ra\cdots |z_{\sigma(N)}\ra ,
	\ee
	where $|z_k\ra=\frac{1}{\sqrt{2}}(e^{\frac{i\delta}{N}}|0\ra+\omega^k|1\ra)$ in which $\omega$ is the $N-$th root of unity, $\omega^N=1$ and $|0\ra$ and $|1\ra$ are the $\sigma_z$ eigenstates in Alice's frame.\\
	
	 \begin{figure}[!ht]
		\centering
		\includegraphics[width=1.3\linewidth]{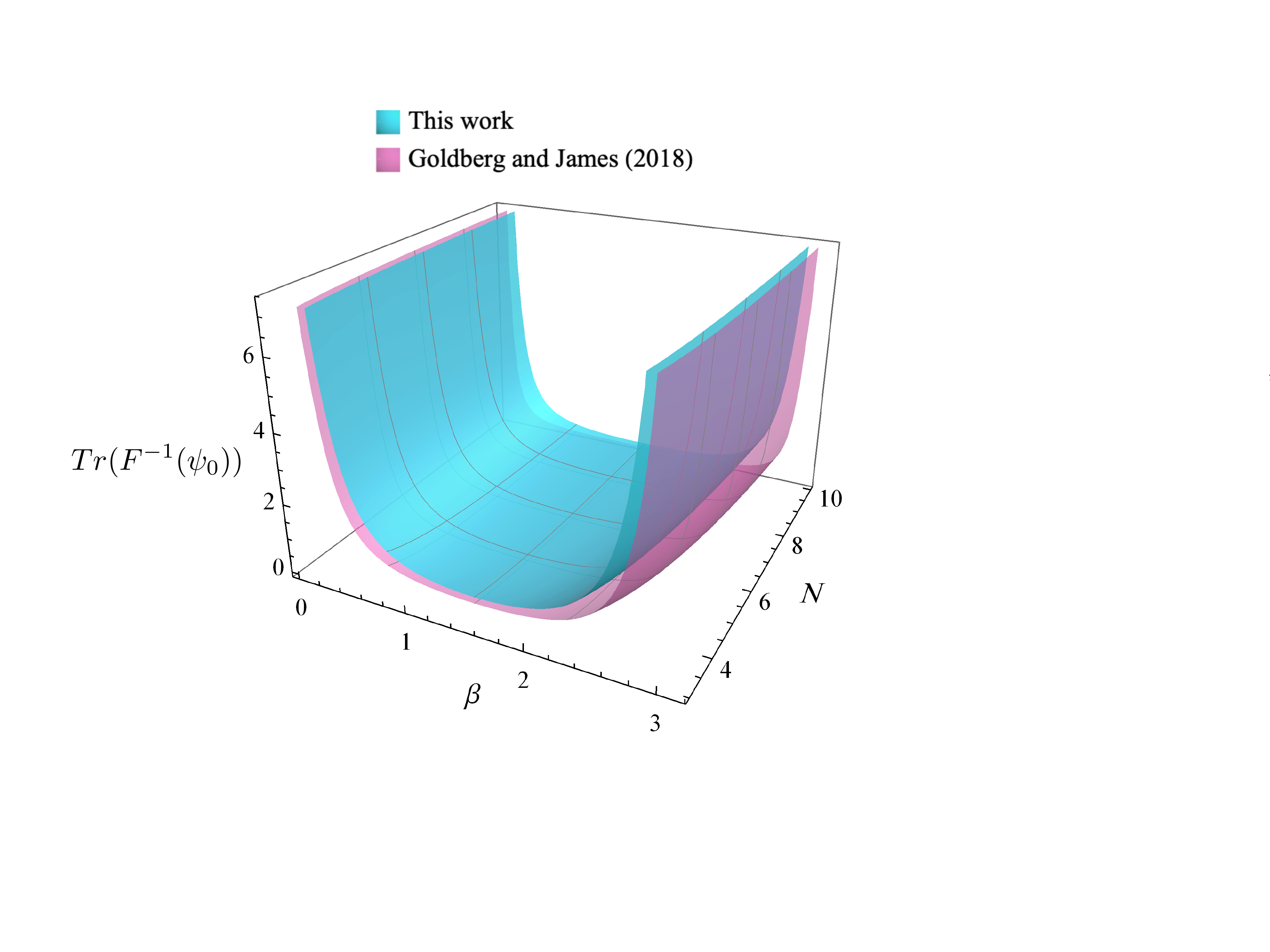}
		\vspace{-3cm}
		\caption{ (Color online) $Tr\big[F^{-1}(U(\a,\b,\gamma)|\psi_0)\big]$ for the case where $|\psi_0\ra$ is the $GHZ$ state (\ref{ghzn0}) in our approach versus the case where $|\psi_0\ra$ is the anti-coherent symmetric state of \cite{gold}. Both depend only on the angle $\beta$ and it is seen that they differ only slightly for almost all values of the rotation angle $\beta$. See the  remark (\ref{re}) for more details.}  
		\label{comp}
	\end{figure}
	
\noindent
\begin{remark}\label{re}
	Although by minimizing $\frac{1}{Tr(\overline{F})}$, (averaged over all rotations), we have shown that the GHZ state  (\ref{ghzn0}) is near optimal, we can calculate $Tr({F^{-1}}(\psi_0))$ for this near optimal state for any specific rotation $U(\a,\beta, \gamma)$, to see how this state works for an arbitrary rotation between the frames. Note that while optimizing  $Tr(\overline{F^{-1}})$, averaged over all rotations, is intractable analytically, it is possible to calculate $Tr({F^{-1}}(|GHZ\ra)$ for arbitrary rotations. 
\end{remark}	
 To this end, we insert the optimal state (\ref{ghzn0}), into the expressions (\ref{fij}) and after straightforward calculations obtain the $F$ matrix 
	 
	 \be
	 F(\psi(\a,\b,\gamma))=	\left(\begin{array}{ccc}N^2\cos^2\beta + N\sin^2\beta & 0 & N^2\cos\beta\\ 0 & N & 0 \\ N^2\cos\beta &0& N^2\end{array}\right),
	 \ee
	 with 
	 
	 \be\label{ttF}
	 \tr(F)=N(1+\sin^2\beta)+N^2(1+\cos^2\beta). 
	 \ee

\noindent Then the inverse of the F matrix becomes
	\be\label{ff-1}
F^{-1}(\psi(\a,\b,\gamma)):=\frac{1}{N\sin^2\beta}\left(\begin{array}{ccc}1 & 0 & -\cos 
		\beta\\ 0 & \sin^2\beta & 0 \\ -\cos\beta &0&  \cos^2\beta + \frac{1}{N}\sin^2\beta\end{array}\right).
\ee

Therefore we find for the GHZ state
\noindent with
\be\label{ff-2}
Tr\big[F^{-1}(U(\a,\b,\gamma)|GHZ\ra)\big]=\frac{1}{N^2}+\frac{2}{N\sin^2\beta}.
\ee
This is to be compared with the result of \cite{gold} which in our notation reads
\be\label{ff-2}
Tr\big[F^{-1}(U(\a,\b,\gamma)|\Phi\ra)\big]=\frac{3}{N(N+1)}\big(1+\frac{2}{\sin^2\beta}\big),
\ee
where $|\Phi\ra$ is the anti-coherent symmetric state of \cite{gold}.
It is seen that both quantities  depends only on the parameter $\beta$ of the rotation, which is the angle between the axes $z$ and $z'$ and is insensitive (symmetric) to the other two rotations. The comparison is done in figure (\ref{comp}). The vertical axis shows $Tr\big[ F^{-1}(\psi(\a,\b,\gamma))\big]$ 
for our near optimal state $\psi_0$ (blue color) and for the anti-coherent polyhedron symmetric state of \cite{gold}(pink color). This figure shows that the lower bound for both these states are very close for all values of $N$ and $\beta$. We should emphasize however that none of these states are the real optimal state which should be found by rigorously minimizing the right hand side of Cram\'{e}r Rao inequality (\ref{crbav}), i.e. minimizing $Tr(\overline{F}^{-1})$. In our case, the shortcoming is that we really maximize  $Tr(\overline{F})$. In case of Goldberg and James, the shortcoming is that for the Euler parameterization of rotations, the basic condition (\ref{cond}) which is true for infinitesimal rotations and hence symmetric parameterization is no longer valid. At a fundamental level, as we have explained in the remark after equation (\ref{crbavin}), for frame alignment in the global approach, when we integrate over all angles of rotations,  no state can be absolutely optimal in the sense of saturating the lower bound of the Cram\'{e}r-Rao bound. What we have obtained with our near-optimal state is the simplicity in experimental realization.

\section{The effect of noise}
It is quite conceivable that the optimal state which has been taken to be a pure state up to now, is affected by noise either by imperfect preparation in the first lab or during transmission between the two labs. The optimal state then changes to a mixed state and it is then desirable to see how the Quantum Fisher Information (QFI) of this optimal state is affected and whether or not it is still useful for estimation of parameters encoded into it. This is the problem which we now study. To this end, we should resort to a formula for Quantum Fisher Information which is appropriate for mixed states and is a generalization of (\ref{eq:QFIpure}). 
According to \cite{Liu2015,Liu2014}, QFI of a mixed state with spectral decomposition $\sum_k \lambda_k \ketbra{\phi_k}{\phi_k}$ (evolving unitarily) is
\begin{equation}\label{eq:mixedQFI}
\rmF_{m,n} = \sum_k  4\lambda_{k}\mathrm{cov}_{k}\left(H_{m},H_{n}\right)-\sum_{k\neq l}\frac{8\lambda_{k}\lambda_{l}}{\lambda_{k}+\lambda_{l}}\mathrm{Re}\left(\langle\phi_{k}|H_{m}|\phi_{l}\rangle\langle\phi_{l}|H_{n}|\phi_{k}\rangle\right).
\end{equation}
For a pure state, this will reduce to (\ref{eq:QFIpure}).
We use the above result and determine the effect of two very important and common sources of noise, namely the dephasing and the depolarizing noise, on the QFI of the optimal state that we already found (\ref{ghzn0}). 

\subsection{Dephasing noise}
Here we study the effect of a very common noise on the optimal state and its viability as a carrier of reference frame information. We consider random phase kicks on all the qubits. This noise changes the optimal state $|\psi_0\ra$ to a mixed state in the form 
\be
|\psi_0\ra\lo \rho=\int dU\ U|\psi_0\ra\la \psi_0|U^{\dagger},
\ee
where 
   $$U=e^{i\theta_1\sigma_{z}}\otimes\cdots \otimes e^{i\theta_N\sigma_{z}},$$
and  $dU=P(\theta_1,\cdots \theta_N)d\theta_1\cdots d\theta_N$, where we have not made any assumption on the probability distribution $P(\theta_1,\cdots \theta_N)$. The random kicks can be correlated or not and can be different on various qubits, the net effect is that they transform the optimal state to 
\be\label{rho-00}
\rho^{dephased}=p|\psi_0\ra\la \psi_0|+\frac{1-p}{2}\big(|\uparrow\cdot\cdot \uparrow\ra\la \uparrow\cdot \cdot \uparrow|+|\downarrow\cdot \cdot \downarrow\ra\la \downarrow\cdot \cdot \downarrow |\big),
\ee
\noindent in which 
\be
p=\int d\theta_1\cdots d\theta_N e^{2i(\theta_1+\cdots+\theta_N)}P(\theta_1,\cdots,\theta_N).
\ee
This means that the optimal state $|\psi_0\ra$  may decohere to product states with probability $1-p$. 
 Note that when there is no phase kick, $p=1$ and we get back our original pure optimal state and when the kicks are completely random, $p=0$, the state becomes a statistical mixture of all-up and all-down spins. Interestingly this state can be written as \cite{shima}
\be\label{rho-0}
\rho^{dephased}=\frac{1+p}{2}|\psi_0\ra\la \psi_0|+\frac{1-p}{2}|\psi'_0\ra\la \psi'_0|,
\ee
where 
\begin{equation}\label{ghzn}
| \psi'_0 \rangle =\frac{|\frac{N}{2},\frac{N}{2}\rangle -  |\frac{N}{2},-\frac{N}{2}\rangle}{\sqrt{2}},
\end{equation}
The noisy optimal state (\ref{rho-0}) is already in the spectral decomposed form. Inserting this in equation (\ref{eq:mixedQFI}) we find after simplification
\ba\label{dephaseF}
F_{m,m}&=&2(1+p) (\la \psi_0|H_m^2|\psi_0\ra-\la \psi_0|H_m|\psi_0\ra^2)\cr&+&2(1-p) (\la \psi'_0|H_m^2|\psi'_0\ra-\la \psi'_0|H_m|\psi'_0\ra^2)\\
 &-&4(1-p^2) (\la \psi'_0|H_m|\psi_0\ra\la \psi_0|H_m|\psi'_0\ra).
\ea

\noindent It turns out that the first two lines of (\ref{dephaseF}) are equal. This is in fact an example of a result mentioned before that the relative phase between the two terms in the optimal state (\ref{ghzn0}) is not relevant. For the cross terms in  the last line we need
\ba
\la \psi'_0|H_a|\psi_0\ra&=&\la \psi'_0|-\cos\beta S_z +\sin\beta (\cos \gamma S_x-\sin\gamma S_y)|\psi_0\ra\cr
&=& 
\cos\beta \la \psi'_0| S_z|\psi_0\ra=\frac{-N}{2}\cos\beta.
\ea
Similarly we find 
\be
\la \psi'_0|H_\beta |\psi_0\ra=\la \psi'_0|-\cos\gamma S_y -\sin\gamma S_x|\psi_0\ra=0,
\ee
where for definiteness we have taken $N>1$ and 
\be
\la \psi'_0|H_\gamma |\psi_0\ra=\la \psi'_0|S_z|\psi_0\ra=\frac{N}{2}.
\ee

\noindent Combining the above, we find
\be
\tr(F)(\rho^{dephased})=\tr(F)(\psi_0)-N^2(1-p^2)(1+\cos^2\beta).
\ee
To compare it with the case for optimal pure state, it is instructive to write their explicit form here, from (\ref{ttF})

\ba
\tr(F)(|\psi_0\ra)&=&N(1+\sin^2\beta)+N^2(1+\cos^2\beta), \cr 
\tr(F)(\rho^{dephased})&=&N(1+\sin^2\beta)+p^2N^2(1+\cos^2\beta). 
\ea
Thus the effect is second order in $p$ but increases also quadratically with the number of particles $N$. 

\subsection{Depolarizing noise}
Another common noise is when the GHZ state undergoes a global depolarizing noise (say in the production process) where the optimal state is changed as 

\be\label{depol}
|\psi_0\ra\lo \rho^{depol}=p|\psi_0\ra\la \psi_0|+\frac{1-p}{2^N}I.
\ee
Incidentally this simple form still holds when each of the qubits locally  undergo depolarizing noise when $N=2$. This is also already in the form of spectral decomposition, with eigenvalues and eigenvectors

\begin{equation}
\begin{cases}
|\psi_0\ra & \xi:=p+\frac{1-p}{2^N},    \\
|\psi_k^\perp\ra & \eta:=\frac{(1-p)}{2^N},
\end{cases}
\label{eqq}
\end{equation}
\noindent where $|\psi_k^\perp\ra$ are $2^N-1$ states which are perpendicular to $|\psi_0\ra$. The simple form of the spectrum and the high degeneracy of the eigenvalue $\eta$, allows us to write (\ref{eq:mixedQFI}) in a compact and simple form. First we rewrite the diagonal elements of $F$ from (\ref{eq:mixedQFI}) as

\begin{equation}\label{eq:mixedQFId}
\rmF_{m,m} = \sum_k  4\lambda_{k}\la \phi_k|H_m^2|\phi_k\ra-\sum_{k, l}\frac{8\lambda_{k}\lambda_{l}}{\lambda_{k}+\lambda_{l}}\left(\langle\phi_{k}|H_{m}|\phi_{l}\rangle\langle\phi_{l}|H_{m}|\phi_{k}\rangle\right).
\end{equation}
Then we use the completeness relation 
$|\psi_0\ra\la \psi_0|+\sum_{k}|\psi_k^\perp\ra\la \psi_k^\perp|=I$ to write for any operator $A$, 
$
\sum_k \la\psi_k^\perp|A|\psi_k^\perp\ra=\tr(A)-\la\psi_0|A|\psi_0\ra 
$
and simplify all the terms, specially the cross terms  in (\ref{eq:mixedQFId}).
 After straightforward calculations we find
\be
F_{m,m}(\rho^{depol})=(\xi+\eta-\frac{4\xi\eta}{\xi+\eta})F_{m,m}(\psi_0),
\ee
or inserting the values of $\xi$ and $\eta$ from (\ref{eqq})
\be
\tr(F)(\rho^{depol})=\Lambda(p,N)\tr F(|\psi_0\ra), \ee  where
 \be
\Lambda(p,N):=p+O(\frac{1-p}{2^N}),
\ee
 \\
showing that the $\tr(F)$ decreases linearly with increase of the level of noise.

\section{Conclusion}
The question we have asked is the following: Given a state $|\psi_0\ra$ and a unitary operator $U(R)$ representing a finite rotation $R(\a,\b,\gamma)$, where $\a,\beta,$ and $\gamma$ are the Euler angles, which state is the optimal state if we want to estimate the angles by making the best measurements on the state $|\psi(\a,\b,\gamma)\ra:=U(R)|\psi_0\ra$. By calculating the Quantum Fisher Information, we have shown that the optimal state is a $|GHZ_N\ra =\frac{1}{\sqrt{2}}(|\uparrow \cdot\cdot \uparrow\ra+|\downarrow\cdot\cdot\downarrow\ra)$ state and have determined in closed form the dependence of the QFI on the Euler angles and also on $N$. We have also determined, again in closed form, the effect of two very common sources of noise on the QFI, if they happen to affect the optimal pure state.  Our work raises several questions. First of all, our work  provides only a near-optimal state which lower bounds a reasonable figure of merit for estimation of the rotation parameters. Considerations of Quantum Fisher information do not normally yield specific measurement procedures for achieving this lower bound. Therefore it will be interesting to find a particular measurement setting for achieving or coming close to any of these lower bounds. Second, it is very interesting to generalize this to a more general estimation problem, i.e. when the parameters of a certain group element $g$ of a continuous group $G$ is to be estimated. In this way connections with the asymmetry properties of states under general group transformations \cite{MarvianSpekkens4} seems to be relevant. Finally we have here considered the case where the two players have no information about the relative orientation of their frames, hence we have used a Haar measure over the rotation group. In \cite{kolen}, it is assumed that the two frames are slightly misaligned, leading to an optimal state which is  symmetric over the Bloch sphere. It will be interesting to see how the optimal state depends on the distribution function, i.e. to the width of a spherical Gaussian distribution of what is known in the literature as a von Mises-Fisher distribution \cite{ahmadi1}.

	%
	\appendix

		\section*{Appendix: Deriving $\tr(\overline{F})$ of Euler rotation}\label{ap-sec:mtqfiEuler}
	To calculate the mean trace QFI $\tr(\overline{F})$ for Euler rotation w.r.t. the distribution Eq.~\eqref{eq:distEulerParam}, we should calculate following integral:
	\begin{align}\label{eq:tr(Fbar)}
	\tr(\overline{F}) &= \int_{\gamma=0}^{2\pi}\int_{\beta=0}^{\pi}\int_{\alpha=0}^{2\pi} \tr F(\alpha,\beta,\gamma)\, z(\alpha,\beta,\gamma) \,\rmd\alpha\rmd\beta\rmd\gamma \\
	&= \frac{1}{8 \pi^2} \int_{\gamma=0}^{2\pi}\int_{\beta=0}^{\pi}\int_{\alpha=0}^{2\pi} (F_{\alpha\alpha}+F_{\beta\beta}+F_{\gamma\gamma})  \sin \beta \,\rmd\alpha\rmd\beta\rmd\gamma,
	\end{align}
	where the quantum Fisher matrix, $F$, is defined in Eq.~\eqref{eq:QFIpure}.
		To calculate each of the three integrals in \eqref{eq:tr(Fbar)} we use the expression for $F_{\a,\a}$ from \eqref{eq:H_a} and calculate

	\begin{multline}
	\int_{\gamma=0}^{2\pi}\int_{\beta=0}^{\pi} \big( \langle H_\a \rangle - \langle H_\a \rangle^2 \big) \sin \beta \, \rmd\beta\rmd\gamma = \\
	2\pi \int_{\beta=0}^{\pi} \cos^2\beta \sin\beta \,\rmd\beta\,\big( \langle S_z^2 \rangle - \langle S_z \rangle^2 \big) \\
	+ \int_{\beta=0}^{\pi} \sin^3\beta \,\rmd\beta\, \int_{\gamma=0}^{2\pi}\cos^2\gamma \,\rmd\gamma\,\big(\langle S_x^2 \rangle-\langle S_x \rangle^2 \big)\\
	+ \int_{\beta=0}^{\pi} \sin^3\beta \,\rmd\beta\, \int_{\gamma=0}^{2\pi}\sin^2\gamma \,\rmd\gamma\,\big(\langle S_y^2 \rangle-\langle S_y \rangle^2 \big) \\
	= \frac{4 \pi}{3} \big( \langle S_x^2 \rangle -\langle S_x \rangle^2 + \langle S_y^2 \rangle -\langle S_y \rangle^2 + \langle S_z^2 \rangle - \langle S_z \rangle^2\big).
	\end{multline}
	Then we use the expression for $F_{\beta,\beta}$ from \eqref{eq:H_beta} and calculate

		\begin{multline}
	\int_{\gamma=0}^{2\pi} \big( \langle H_\beta^2\rangle - \langle H_\beta \rangle^2\big)\,\rmd\gamma = \\
	\int_{\gamma=0}^{2\pi} \cos^2\gamma\,\rmd\gamma \big( \langle S^2_y \rangle - \langle S_y \rangle^2\big) + \int_{\gamma=0}^{2\pi} \sin^2\gamma\,\rmd\gamma \big( \langle S^2_x \rangle - \langle S_x \rangle^2\big) \\
	= \pi \big( \langle S^2_y \rangle - \langle S_y \rangle^2 + \langle S^2_x \rangle - \langle S_x \rangle^2\big).
	\end{multline}
\noindent	Finally we the expression for $F_{\gamma,\gamma}$ from \eqref{eq:H_gamma} and calculate 
	\be
	\int_{\gamma=0}^{2\pi} \big( \langle H_\gamma^2\rangle - \langle H_\gamma \rangle^2\big)\,\rmd\gamma = \langle S_z^2\rangle - \langle S_z \rangle^2 .
	\ee

\noindent	Putting everything together the final expression for $\tr(\overline{F})$ is simplified to 
	\begin{equation}
	\tr(\overline{F}) = \frac{10}{3} \big( \langle S_x^2 \rangle -\langle S_x \rangle^2 \big) + \frac{10}{3} \big( \langle S_y^2 \rangle -\langle S_y \rangle^2 \big) + \frac{16}{3} \big( \langle S_z^2 \rangle - \langle S_z \rangle^2\big),
	\end{equation}
	which is the expression used in \eqref{eq:mtqEuler}.


\begin{thebibliography}{}
	\bibitem{gio} Vittorio Giovannetti, Seth Lloyd, and Lorenzo Maccone, \emph{Quantum metrology},	
	Phys. Rev. Lett. 96, 010401 (2006).

\bibitem{dariano}GM D'Ariano, PL Presti, MGA Paris, Using entanglement improves the precision of quantum measurements, Physical review letters 87 (27), 270404.

\bibitem{paris} Matteo G. A. Paris, \emph{Quantum estimation for quantum technology
MGA Paris}, International Journal of Quantum Information 7 (supp01), 125-137.

\bibitem{paris2}GM D'Ariano, MGA Paris, MF Sacchi, Quantum tomography
Advances in Imaging and Electron Physics 128, 206-309
	
	\bibitem{gispop} N. Gisin and S. Popescu, \emph{Spin Flips and Quantum Information for Antiparallel Spins}, Phys. Rev. Lett. \textbf{83}, 432 (1999).
	
	
	\bibitem{mass} S. Massar and S. Popescu, \emph{Optimal Extraction of Information from Finite Quantum Ensembles}, Phys. Rev. Lett. \textbf{74}, 1259 (1995).
	
	
\bibitem{peres1} Asher Peres and Petra F. Scudo, \emph{Entangled Quantum States as Direction Indicators}, Phys. Rev. Lett. 86 (2001) 4160.

\bibitem{peres2} Asher Peres and Petra F. Scudo, \emph{Transmission of a Cartesian Frame by a Quantum System}, Phys. Rev. Lett. 87 (2001) 167901.

	
\bibitem{reza} F. Rezazadeh, A. Mani, V. Karimipour, \emph{Secure alignment of coordinate systems by using quantum correlation}, Physical Review A 96 (2), 022310.

\bibitem{bagan} E. Bagan, M. Baig, and R. Mun\~{o}z Tapia, \emph{Aligning Reference Frames with Quantum States}, Phys. Rev. Lett. 87, 257903 (2001).

\bibitem{chir}G. Chiribella, G. M. D'Ariano, P. Perinotti  and M. E. Sacchi, \emph{Efficient Use of Quantum Resources for the Transmission of a Reference Frame}, Phys. Rev. Lett.  {\bf 93}, 180503 (2004).
	
\bibitem{kolen} P. Kolenderski and R. Demkowicz-Dobrzanski, \emph{Optimal state for keeping reference frames aligned and the platonic solids}, Phys. Rev. A 78, 052333 (2008).

\bibitem{gold} A. Z. Goldberg and D. F. V. James, \emph{Quantum-limited Euler angle measurements using anticoherent states} Phys. Rev. A 98, 032113 (2018).	

\bibitem{Rudolph1} Stephen D. Bartlett, Terry Rudolph, R. W. Spekkens, \emph{Classical and quantum communication without a shared reference frame}, Phys. Rev. Lett. 91, 027901 (2003).

\bibitem{Rudolph2} Stephen D. Bartlett, Terry Rudolph, Robert W. Spekkens, Peter S. Turner, \emph{Quantum communication using a bounded-size quantum reference frame}, New J. Phys. 11, 063013 (2009).

\bibitem{rezaqkd} F Rezazadeh, A Mani, V Karimipour, \emph{Quantum Key distribution with no shared reference frame},  Quant. Info. Processing, {\bf 9}, 104, (2019).Phys. Rev. A 96, 022310 (2017).
	
	
	\bibitem{Beheshti} Ali Beheshti, Sadegh Raeisi, Vahid Karimipour, \emph{Entanglement-assisted communication in the absence of shared reference frame}, Physical Review A 99 (4), 042330.

\bibitem{ahmadi1}Mehdi Ahmadi, Alexander R. H. Smith, Andrzej Dragan, \emph{Communication between inertial observers with partially correlated reference frames}, Phys. Rev. A 92, 062319 (2015).

\bibitem{ahmadi2} Dominik Šafránek, Mehdi Ahmadi and Ivette Fuentes, \emph{Quantum parameter estimation with imperfect reference frames, New Journal of Physics}, 
New J. Phys. 17 (2015) 033012.

\bibitem{ahmadi3}M Ahmadi, D Jennings, T Rudolph, \emph{The Wigner–Araki–Yanase theorem and the quantum resource theory of asymmetry}, New Journal of Physics 15 (1), 013057.

\bibitem{MarvianSpekkens1} Iman Marvian, Robert W. Spekkens, \emph{The asymmetry properties of pure quantum states}, Phys. Rev. A 90, 014102 (2014).

\bibitem{MarvianSpekkens2} Iman Marvian, Robert W. Spekkens, \emph{A no-broadcasting theorem for quantum asymmetry and coherence and a trade-off relation for approximate broadcasting}, Phys. Rev. Lett. 123, 020404.

\bibitem{MarvianSpekkens3} Iman Marvian, Robert W. Spekkens, \emph{How to quantify coherence: Distinguishing speakable and unspeakable notions}, Phys. Rev. A 94, 052324 (2016).


\bibitem{MarvianSpekkens4}Iman Marvian and  Robert W. Spekkens, \emph{
The theory of manipulations of pure state asymmetry: I. Basic tools, equivalence classes and single copy transformations}
New Journal of Physics, {\bf 15} 033001, (2013).


\bibitem{zimba} J. Zimba, \emph{“Anticoherent” Spin States via the Majorana Representation}, Electronic Journal of Theoretical Physics 3, 143 (2006).

\bibitem{maj}E. Majorana, \emph{Oriented atoms in a variable magnetic field}, Nuovo Cimento {\bf 9}, 43 (1932).

	
		\bibitem{ghzexp1} Manuel Erhard, Mehul Malik, Mario Krenn, Anton Zeilinger, \emph{Experimental Greenberger Horne Zeilinger Entanglement beyond Qubits}, Nature Photonics 12, 759-764 (2018).
	
	\bibitem{ghzexp2} D. Bouwmeester, J.-W. Pan, M. Daniell, H. Weinfurter, and A. Zeilinger, \emph{Observation of Three-Photon Greenberger Horne Zeilinger Entanglement} Physical Review Letters 82, 1345 (1999).
	
	\bibitem{ghzexp3} P. Jian-Wei, D. Bouwmeester, M. Daniell, H. Weinfurter, and A. Zeilinger, \emph{Experimental test of quantum nonlocality in three-photon Greenberger Horne Zeilinger entanglement}, Nature 403, 515 (2000).
	
	\bibitem{ghzexp4} X.-L. Wang, L.-K. Chen, W. Li, H.-L. Huang, C. Liu, C. Chen, Y.-H. Luo, Z.-E. Su, D. Wu, Z.-D. Li, et al., \emph{Experimental Ten-Photon Entanglement}, Phys. Rev. Lett. 117, 210502 (2016).




\bibitem{cr1} H. Cram\'{e}r, \emph{Mathematical Methods of Statistics}, Princeton, N.J. :Princeton Univ. Press. (1946).
\bibitem{cr2} C. R.  Rao, \emph{Information and the accuracy attainable in the estimation of statistical parameters}, Bulletin of the Calcutta Mathematical Society. 37: 81–89. MR 0015748 (1945).

		
\bibitem{hels} Carl W. Helstrom, \emph{Quantum detection and estimation theory}, Academic Press, 1976.

\bibitem{petz} D\'{e}nes Petz, \emph{Quantum Information Theory and Quantum Statistics}, Theoretical and Mathematical Physics, Springer-Verlag Berlin Heidelberg, 2008.


		

		

\bibitem{matsu}K. Matsumoto, \emph{A new approach to the Cramer-Rao type bound of the pure state model}, J. Phs. A {\bf 35}, 3111 (2002).
	\bibitem{gill1} 	R.D. Gill and  B.Y. Levit, \emph{Applications of the van Trees inequality: a Bayesian Cramér-Rao bound}, 	Bernoulli 1 (1-2), 59-79 (1995)
	\bibitem{gill2} 	R. D. Gill and  S. Massar, \emph{State estimation for large ensembles}
	Physical Review A 61, 042312 (2000).
	
	
\bibitem{conv} Kenneth Nordstr\"{o}m, \emph{Convexity of the inverse and Moore–Penrose inverse}, Linear Algebra and its Applications, Vol. 434 (6), pp. 1489-1512, 2011.

\bibitem{Liu2015} Liu, J., Jing, X., and Wang, X., \emph{Quantum metrology with unitary parametrization processes}, Sci Rep 5, 8565 (2015); DOI: 10.1038/srep08565.

\bibitem{Liu2014} Liu, J., Xiong, H.-N., Song, F., and Wang, X., \emph{Fidelity susceptibility and quantum Fisher information for density operators with arbitrary ranks}, Phyica A 410, 167 (2014); DOI: 10.1016/j.physa.2014.05.028.

	\bibitem{BellIneq} P. Shadbolt, et al., \emph{Guaranteed violation of a Bell inequality without aligned reference frames or calibrated devices}, Scientific Reports 2, 470 (2012).	

\bibitem{shima} Shima Emamipanah, Marzieh Asoudeh, and Vahid Karimipour, \emph{Entangled States as Robust and Re-usable Carriers of Information},  Quantum Information Processing, {\bf 19}, 357 (2020).







	\end{thebibliography}
\end{document}